\documentclass[aps,prb,reprint,preprintnumbers,superscriptaddress,amsmath,amssymb,bibnotes,longbibliography]{revtex4-2}

\usepackage{graphicx}
\usepackage{dcolumn}
\usepackage{bm}
\usepackage[colorlinks,linkcolor=blue,anchorcolor=blue,citecolor=blue,urlcolor=blue,filecolor=blue,menucolor=blue,runcolor=blue]{hyperref}
\usepackage{ulem}

\begin{document}
\title{Magnetic properties of the layered heavy fermion antiferromagnet CePdGa$_6$}
\author{H. Q. Ye}
\affiliation{Center for Correlated Matter and Department of Physics,
Zhejiang University, Hangzhou 310058, China}
\author{T. Le}
\affiliation{Center for Correlated Matter and Department of Physics,
Zhejiang University, Hangzhou 310058, China}
\author{H. Su}
\affiliation{Center for Correlated Matter and Department of Physics,
Zhejiang University, Hangzhou 310058, China}
\author{Y. N. Zhang}
\affiliation{Center for Correlated Matter and Department of Physics,
Zhejiang University, Hangzhou 310058, China}
\author{S. S. Luo}
\affiliation{Center for Correlated Matter and Department of Physics,
Zhejiang University, Hangzhou 310058, China}
\author{M. J. Gutmann}
\affiliation{ISIS Facility, Rutherford Appleton Laboratory, Chilton, Didcot Oxon OX11 0QX, United Kingdom}
\author{H. Q. Yuan}
\affiliation{Center for Correlated Matter and Department of Physics,
Zhejiang University, Hangzhou 310058, China}
\affiliation  {Zhejiang Province Key Laboratory of Quantum Technology and Device, Department of Physics, Zhejiang University, Hangzhou 310058, China}
\affiliation  {State Key Laboratory of Silicon Materials, Zhejiang University, Hangzhou 310058, China}
\affiliation{Collaborative Innovation Center of Advanced
Microstructures, Nanjing University, Nanjing 210093, China}
\author{M. Smidman}
\email{msmidman@zju.edu.cn}
\affiliation{Center for Correlated Matter and Department of Physics,
Zhejiang University, Hangzhou 310058, China}
\affiliation  {Zhejiang Province Key Laboratory of Quantum Technology and Device, Department of Physics, Zhejiang University, Hangzhou 310058, China}
\date{\today}
\addcontentsline{toc}{chapter}{Abstract}

\begin{abstract}

We report the magnetic properties of the layered heavy fermion antiferromagnet CePdGa$_{6}$, and their evolution upon tuning with the application of magnetic field and pressure. CePdGa$_{6}$ orders antiferromagnetically below $T\rm_{N}$ = 5.2~K, where there is evidence for heavy fermion behavior from an enhanced Sommerfeld coefficient. Our results are best explained by a magnetic ground state of ferromagnetically coupled layers of Ce $4f$-moments orientated along the $c$-axis, with antiferromagnetic coupling between layers. At low temperatures we observe two metamagnetic transitions  for fields applied along the $c$-axis corresponding to spin-flip transitions, where the lower transition is to a different magnetic phase with a magnetization one-third of the saturated value. From our analysis of the magnetic susceptibility, we propose a CEF level scheme which accounts for the Ising anisotropy at low temperatures, and we find that the evolution of the magnetic ground state can be explained considering both antiferromagnetic exchange between nearest neighbor and next nearest neighbor layers, indicating the influence of long-range  interactions. Meanwhile we find little change of $T\rm_{N}$ upon applying hydrostatic pressures up to 2.2~GPa, suggesting that significantly higher pressures are required to examine for possible quantum critical behaviors.

\end{abstract}

\maketitle

\section{Introduction}

Heavy fermion compounds are prototypical examples of strongly correlated electron systems, and have been found to host a range of emergent phenomena including unconventional superconductivity, complex magnetic order and strange metal behavior \cite{weng2016multiple,si2010heavy,coleman2007heavy}. Ce-based heavy fermions contain a Kondo lattice of Ce-ions with an unpaired $4f$ electron, which can both couple to other $4f$ moments via the Ruderman-Kittel-Kasuya-Yosida (RKKY) interaction and undergo the Kondo interaction due to hybridization with the conduction electrons. Here the RKKY interaction gives rise to long-range magnetic order, while the Kondo interaction favors a non-magnetic Fermi-liquid ground  state with greatly enhanced quasiparticle masses. Due to the small energy scales, the relative strengths of these competing interactions can often be tuned by non-thermal parameters such as pressure, magnetic fields and chemical doping \cite{Doniach_1977}, and in many cases the magnetic ordering can be continuously suppressed to zero temperature at a quantum critical point (QCP).

A major question for heavy fermion systems is the relationship between quantum criticality, and the dome of unconventional superconductivity sometimes found to encompass the QCP. CeIn$_3$ is a canonical example of this phenomenon, which at ambient pressure orders antiferromagnetically below $T_{\rm N}$ = 10.1~K, but exhibits a pressure-induced QCP around 2.6~GPa, which is surrounded by a superconducting dome with a maximum $T_{\rm c}$ of 0.2~K \cite{mathur1998magnetically}. The layered Ce$M$In$_5$ ($M$= transition metal) compounds consist of alternating layers of $M$In$_2$ and CeIn$_3$ along the $c$-axis \cite{Thompson_2012}, and among the remarkable properties is a significantly enhanced superconducting $T_{\rm c}$ for the $M$= Rh and Co systems, reaching over 2~K \cite{Tuson_2006,Petrovic_2001}, giving a strong indication that quasi-two-dimensionality is important for promoting heavy fermion superconductivity. Meanwhile the Ce$_2M$In$_8$ compounds correspond to a stacked arrangement of two units of CeIn$_3$, and one of $M$In$_2$ \cite{PhysRevB.64.144411}, and  are expected to have an intermediate degree of two dimensionality relative to  Ce$M$In$_5$. Correspondingly, the superconducting phases have lower $T_{\rm c}$ values  of 0.4 and 0.68~K for Ce$_2$CoIn$_8$ \cite{chen2002observation} and Ce$_2$PdIn$_8$ \cite{PhysRevLett.103.027003} at ambient pressure, and a maximum of $T\rm_c$ = 2~K at 2.3 GPa for  Ce$_2$RhIn$_8$ \cite{PhysRevB.67.020506}. On the other hand, these different series of related Ce-based heavy fermion systems also exhibit different magnetic ground states and crystalline electric field (CEF) level schemes \cite{Curro2000,Bao2000,Bao2001,Christianson2002,Christovam2019} and therefore it is challenging to disentangle the role of these factors from that of the reduced dimensionality. The elucidation of the interplay between these different aspects requires examining additional families of layered Ce-based heavy fermion systems for quantum critical behaviors, as well as detailed characterizations of the magnetic ground states and exchange interactions.

The properties  of layered Ce-based heavy fermion gallides have been less studied than the indium-based systems. CeGa$_6$ has a layered tetragonal structure (space group $P4/nbm$), with four Ga-layers between each Ce layer \cite{Pelleg1981}. This compound orders magnetically below $T_{\rm N}$ = 1.7~K, and there is evidence for the build-up of magnetic correlations at significantly higher temperatures  \cite{Erik_1999}. A more layered structure is realized in the Ce$_2M$Ga$_{12}$  ($M$= Cu, Ni, Rh, Pd, Ir, Pt) series, where the Ce-layers are alternately separated by four Ga-layers, and units of $M$Ga$_6$, leading to a larger interlayer separation of the Ce-atoms  \cite{Macaluso_2005,Cho_2008}. Several members of this series show evidence for both antiferromagnetism and heavy fermion behavior \cite{Macaluso_2005,Cho_2008, Nallamuthu_2014,Sichevych_2012,Gnida_2013,PhysRevB.101.024421}, where pressure can readily suppress the antiferromagnetic transitions of  Ce$_2$NiGa$_{12}$ and Ce$_2$PdGa$_{12}$ \cite{Kawamura_2014,Ohara_2012}, while evidence for field-induced critical fluctuations is revealed in Ce$_2$IrGa$_{12}$  \cite{PhysRevB.101.024421}.

\begin{figure}[h]
\includegraphics[width=8cm]{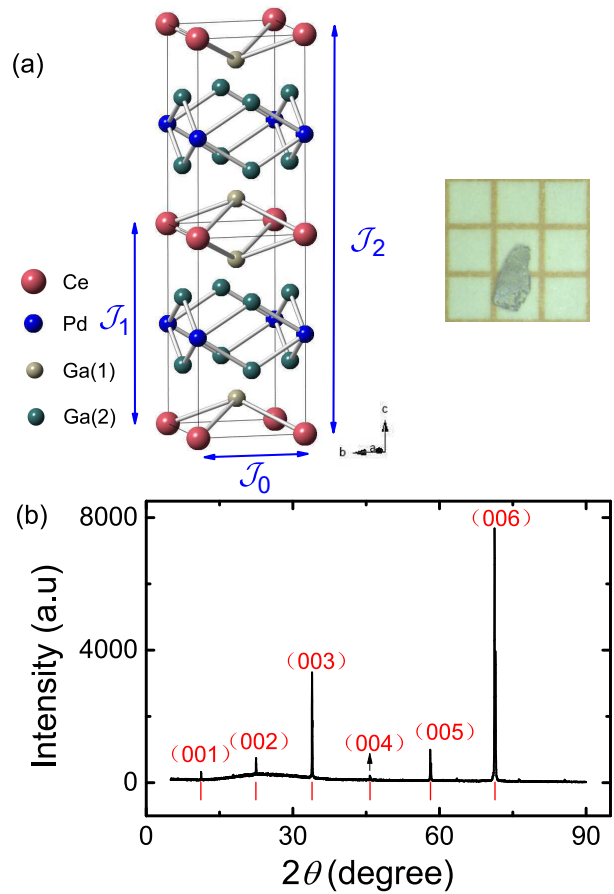}
\setlength{\abovecaptionskip}{-0.cm}   %
\caption{(Color online) (a)  Crystal structure of CePdGa$_{6}$ where the red, blue and green atoms correspond to  Ce, Pd and Ga, respectively. $\mathcal{J}_0$ represents magnetic exchange interactions between nearest neighbor Ce atoms within the $ab$-plane, $\mathcal{J}_1$ is between nearest neighboring layers and $\mathcal{J}_2$ is between next nearest layers. An image of a typical single crystal of CePdGa$_{6}$ is also displayed, where each square in the background is $2$~mm~$\times~2$~mm. (b) X-ray diffraction pattern  measured on a single crystal of CePdGa$_{6}$. The red dashes correspond to the positions of the ($00l$) Bragg peaks, indicating that the [001] direction is perpendicular to the large face of the plate-like samples.
}
\label{fig1}
\end{figure}

CePdGa$_6$ has a different layered tetragonal structure (space group $P4/mmm$) displayed in Fig.~\ref{fig1}(a), consisting of square layers of Ce-atoms, with each Ce contained in a CeGa$_4$ prism, separated by PdGa$_2$ layers \cite{MACALUSO2003296}. Correspondingly, there is a distance between Ce-layers of 7.92~\AA, while the nearest neighbor in-plane Ce-Ce separation is 4.34~\AA, compared to respective values of 7.54~\AA~ and 4.65~\AA~in CeRhIn$_5$ \cite{Hegger_2000}. CePdGa$_6$ orders antiferromagnetically below $T_{\rm N}$ = 5.2~K, and heavy fermion behavior is evidenced by an enhanced Sommerfeld coefficient \cite{MACALUSO2003296,Macaluso_2005}. As such, CePdGa$_6$ is a good candidate to look for novel behaviors arising in quasi-two-dimensional heavy fermion systems, but there is both a lack of  detailed characterizations of the magnetic ground state, and no reports of the evolution under pressure. In addition, most measurements of CePdGa$_6$  are reported in Ref.~\onlinecite{MACALUSO2003296}, where the results are  affected by the inclusion of an extrinsic antiferromagnetic phase Ce$_2$PdGa$_{12}$, which can be eliminated using a modified crystal growth procedure \cite{Macaluso_2005}.

In this article we report detailed measurements of the magnetic properties of single crystals of CePdGa$_6$, including their evolution upon applying magnetic fields and hydrostatic pressure. We find that CePdGa$_6$ orders antiferromagnetically in zero-field, where the Ce-moments are orientated along the $c$-axis and align ferromagnetically within the $ab$-plane, but there is antiferromagnetic coupling between layers. At low temperatures, two metamagnetic transitions are observed for fields along the $c$-axis, the lower of which corresponds to a spin-flip transition to a phase with  magnetization one-third of the saturated value. From our analysis of the magnetic susceptibility, we propose a CEF level scheme which can explain the low temperature Ising anisotropy, and we find that from considering interactions between the nearest-neighbor and next nearest neighbor Ce-layers, the field evolution of the magnetic state can be well accounted for.

\section{Experimental details}

Single crystals of CePdGa$_{6}$ were grown using a Ga self-flux method with a molar ratio of Ce:Pd:Ga of 1:1.5:15 \cite{Macaluso_2005}. Starting materials of Ce ingot (99.9$\%$), Pd powder (99.99$\%$) and Ga pieces (99.99$\%$) were loaded into an alumina crucible which was sealed in an evacuated quartz tube. The tube was heated to 1150$~^\circ$C and held at this temperature for two hours, before being rapidly cooled to 500$~^\circ$C at a rate of 150 K/h  and then cooled more slowly to 400$~^\circ$C at 8 K/h. After being held at 400$~^\circ$C for two weeks, the tube was removed from the furnace,  and  centrifuged to remove excess Ga. The obtained crystals are plate-like with typical dimensions $2\times1.5\times0.3$~mm$^3$. Note that when slower cooling rates of 6 K/h or 4 K/h were used, the resulting crystals were significantly smaller. Single crystals of the non-magnetic analog LaPdGa$_6$ were also obtained using a similar procedure. The composition was confirmed  using a cold field emission scanning electron microscope (SEM) equipped with an  energy dispersive x-ray spectrometer. The phase of the crystals were checked using both a  PANalytical X’Pert MRD powder diffractometer using Cu-K$\alpha$ radiation, and a Rigaku-Oxford diffraction Xtalab synergy single crystal diffractometer equipped with a HyPix hybrid pixel array detector using Mo-K$\alpha$ radiation. The obtained lattice parameters from the single crystal diffraction data of $a$ = 4.3446(3)~$\rm\AA$ and $c$ = 7.9173(10)~$\rm\AA$ are in excellent agreement with previous reports \cite{MACALUSO2003296}. Measurements of a crystal using the powder diffractometer are displayed in Fig.~\ref{fig1}(b), where all the Bragg peaks are well-indexed by the (00$l$) reflections of CePdGa$_6$, demonstrating that the $c$-axis is perpendicular to the large face of the crystals. Resistivity and specific heat measurements were performed in  applied fields up to 14~T using a Quantum Design Physical Property Measurement System (PPMS-14) down to 1.8~K, and to 0.3~K using a $^3$He insert. Resistivity measurements were performed after spot welding four Pt wires to the surface, with the excitation current in the $ab$-plane. Magnetization measurements were performed in the range 1.8 - 300 K in applied fields up to 5~T using a Quantum Design Magnetic Property Measurement System (MPMS) SQUID magnetometer. Heat capacity measurements under pressure were carried out in a piston cylinder cell, using an ac calorimetric method.

\section{results}

\begin{figure}[t]
\includegraphics[width=8.6cm]{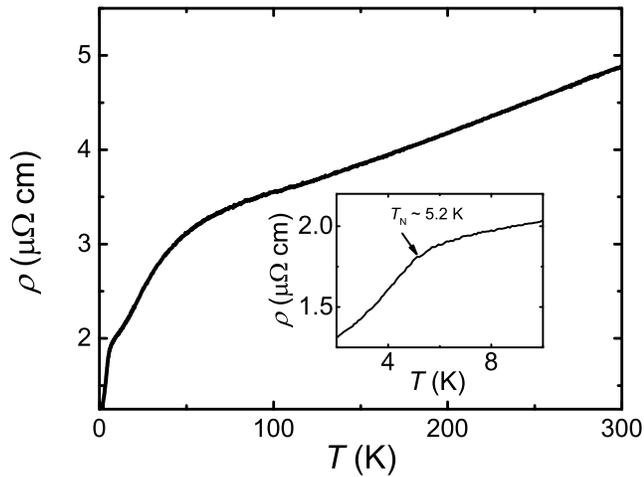}
\caption{(Color online) Temperature dependence of the resistivity $\rho(T)$ of CePdGa$_{6}$ between 1.8 and 300 K. The inset displays the low temperature resistivity, where there is a sharp anomaly at the antiferromagnetic transition.
}
\label{fig2}
\end{figure}

\begin{figure}[t]
\includegraphics[width=8.6cm]{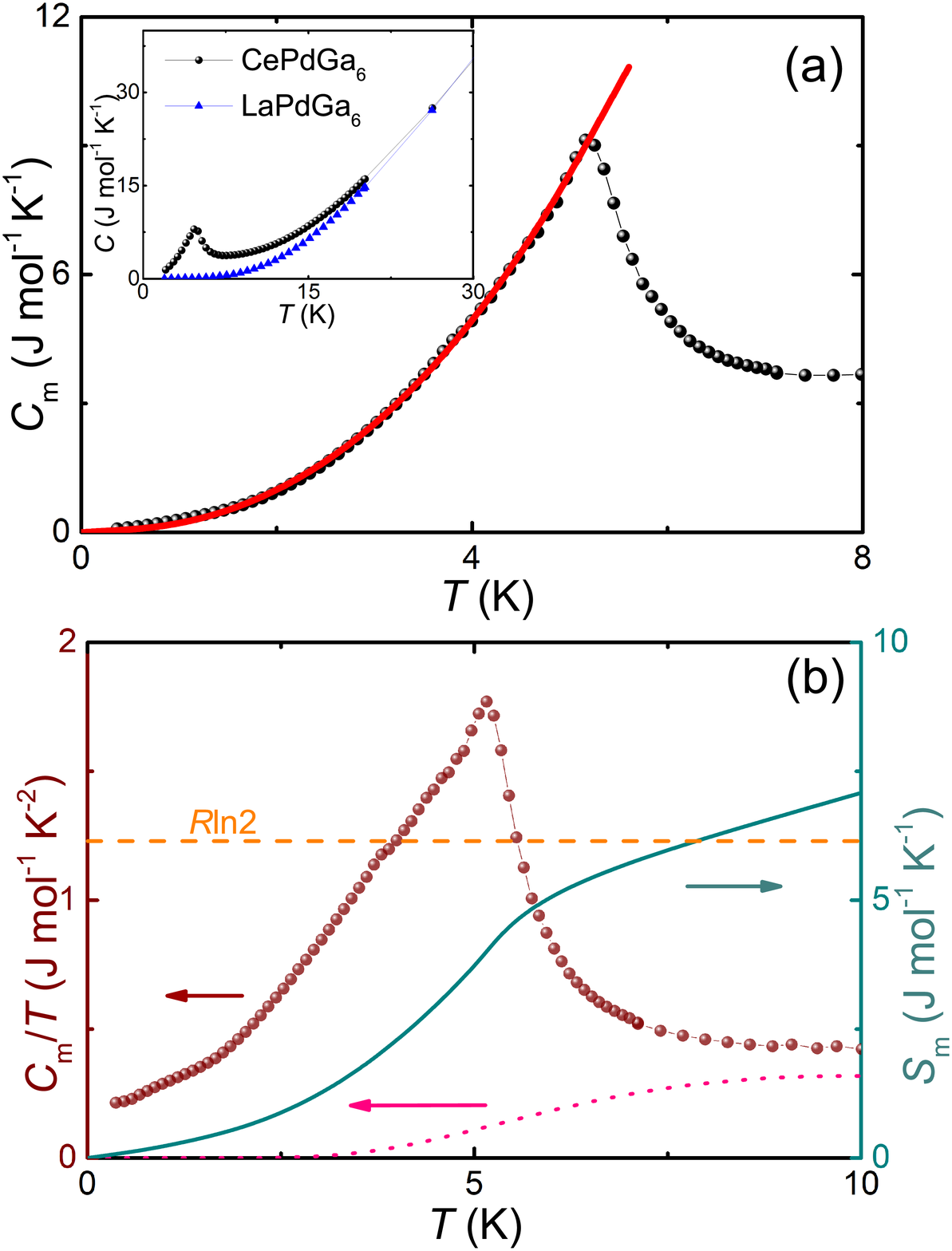}
\caption{(Color online) (a) Magnetic contribution to the specific heat $C\rm_m$ at low temperatures, where the red solid line shows the results from fitting with Eq.~\ref{EqSpinW}. The inset shows the total specific heat $C$ of CePdGa$_{6}$ and the non-magnetic analog LaPdGa$_{6}$. (b) Temperature dependence of $C\rm_m/T$ and the magnetic entropy $S\rm_m$ of CePdGa$_{6}$. The pink dotted line displays the low temperature contribution to the specific heat calculated from the CEF scheme deduced from the analysis of $\chi(T)$.
}
\label{fig3}
\end{figure}

\begin{figure}[h]
\includegraphics[width=8.6cm]{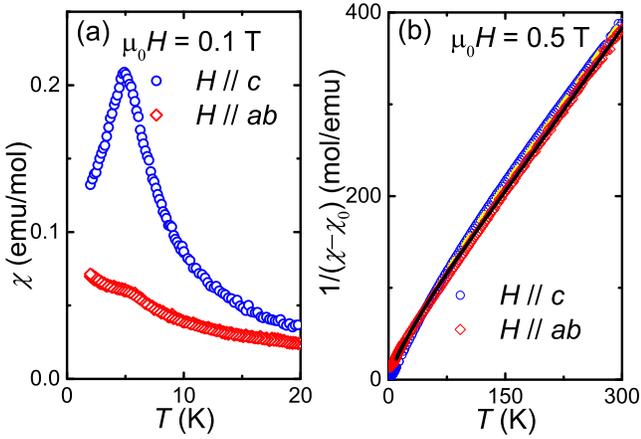}
\caption{(Color online) (a) Low temperature magnetic susceptibility $\chi(T)$ of CePdGa$_{6}$,  with an applied field of $\mu_0H$ = 0.1 T both parallel to the $c$-axis and within the $ab$-plane. (b) Temperature dependence of 1/($\chi$-$\chi_0$) up to 300 K for 0.5~T applied along the  two field directions, where the dashed and solid lines show the results from fitting with the CEF model described in the text.
 }
\label{fig4}
\end{figure}

\subsection{Antiferromagnetic transition and CEF excitations of CePdGa$_{6}$ }

Figure~\ref{fig2} displays the temperature dependence of the resistivity $\rho(T)$ of CePdGa$_{6}$ between 1.8 and 300 K, which has a residual resistivity ratio [RRR = $\rho$(300~K)/$\rho(2~K)]=3.8$. A broad shoulder is observed at around 50 K, which likely arises due to both the Kondo effect, and as a consequence of CEF excitations. At higher temperatures, quasilinear behavior is observed, which could be due to electron-phonon coupling. As shown in the inset, there is an anomaly at around $T_{\rm N}=5.2$~K, below which $\rho(T)$ decreases more rapidly with decreasing temperature, which corresponds to the antiferromagnetic transition reported previously \cite{Macaluso_2005}, while no signature of the spurious transition at higher temperatures is detected \cite{MACALUSO2003296}. The total specific heat of CePdGa$_{6}$ and nonmagnetic  isostructural LaPdGa$_{6}$ are shown in the inset of Fig.~\ref{fig3}(a). The temperature dependence of the magnetic contribution to the specific heat $C_m$ was estimated by subtracting the data of LaPdGa$_{6}$, which is shown in Fig.~\ref{fig3}(a), while the specific heat coefficient $C_{\rm m}/T$ and the magnetic entropy $S_{\rm m}$ of  CePdGa$_{6}$ are displayed in Fig.~\ref{fig3}(b). A pronounced  $\lambda$-like anomaly is observed at $T\rm_{N}$ = 5.2~K, as is typical for a second-order magnetic phase transition. For $T>T_{\rm N}$, $C_{\rm m}/T$ increases with decreasing temperature, and extrapolates to a relatively large zero temperature value of $250~\mathrm{mJ} / \mathrm{mol}~\mathrm{K}^{2}$. As discussed below, the analysis of the magnetic susceptibility $\chi(T)$ suggests the presence of a low lying CEF level, which could contribute to  $C_{\rm m}/T$  in this temperature range. The dotted line in Fig.~\ref{fig3}(b)  shows the calculated  $C_{\rm m}/T$ for the CEF level scheme described below, which has a sizeable value in the vicinity of the transition. Subtracting the contribution from the CEF at $T\rm_{N}$ yields an estimate of $\gamma\sim121.4~\mathrm{mJ} / \mathrm{mol}~\mathrm{K}^{2}$ associated with the ground state doublet, and such an enhanced value could arise both due to heavy fermion behavior, as well as the presence of short range magnetic correlations, as inferred in CeRhIn${_5}$\cite{PhysRevB.69.024419,PhysRevB.66.054433}. The data below $T\rm_N$ were analyzed using \cite{de2000quantum}:
\begin{equation}
\begin{aligned}
C_{\mathrm{m}} &=\gamma T+c \Delta_{\mathrm{SW}}^{7 / 2} \sqrt{T} \exp \left(\frac{-\Delta_{\mathrm{SW}}}{T}\right) \\
& \times\left[1+\frac{39 T}{20 \Delta_{\mathrm{SW}}}+\frac{51}{32}\left(\frac{T}{\Delta_{\mathrm{SW}}}\right)^{2}\right]
\end{aligned}
\label{EqSpinW}
\end{equation}
\noindent where the first term corresponds to the electronic contribution and the second term arises due to antiferromagnetic spin-waves. Here the  coefficient $c$ is related to the spinwave stiffness $D$ via $c \propto D^{-3}$, while $\Delta_{\text {SW }}$ is the spin-wave gap. The results from fitting the zero-field data are displayed in the main panel of  Fig.~\ref{fig3}(a), where $\gamma=121.4~\mathrm{mJ} / \mathrm{mol}~\mathrm{K}^{2}$ was fixed, yielding $\Delta_{\mathrm{SW}}=2.3~\mathrm{K}$ and $c=23~\mathrm{mJ} / \mathrm{mol} ~\mathrm{K}^{2}$. The moderate value of $\Delta_{\text {SW }}$ is smaller than $T_{\rm N}$, unlike the layered heavy fermions gallides Ce$_2$PdGa$_{12}$ and Ce$_2$IrGa$_{12}$ where $\Delta_{\text {SW }}>T_{\rm N}$ \cite{PhysRevB.101.024421,Gnida_2013}, likely reflecting  the weaker magnetocrystalline anisotropy in CePdGa$_6$. The temperature dependence of the magnetic entropy $S_{\mathrm{m}}$  of CePdGa$_6$ is also displayed in Fig.~\ref{fig3}(b), obtained by integrating $C_{\mathrm{m}}/T$, where $C_{\mathrm{m}}/T$ was linearly extrapolated below 0.4~K. At $T_{\mathrm{N}}$, $S_{\mathrm{m}}$ reaches $0.76R \ln 2$, which together with the expected sizeable contribution from the excited CEF level discussed above, suggests a reduced entropy corresponding to the ground state doublet due to Kondo screening.

Figure~\ref{fig4}(a) displays the temperature dependence of the magnetic susceptibility $\chi(T)$ of CePdGa$_{6}$ at low temperatures, with an applied field of $\mu_0H$ = 0.1 T along the $c$-axis and within the $ab$-plane, which both exhibit an anomaly at $T\rm_{N}$. At low temperatures, $\chi(T)$ is significantly larger for fields along the $c$-axis than in the $ab$-plane, demonstrating that the $c$-axis is the easy-axis of magnetization. At $T\rm_{N}$, there is a peak in $\chi(T)$ for $H\parallel c$, while for  $H\parallel ab$ $\chi(T)$ weakly increases below $T\rm_{N}$, indicating that this corresponds to an antiferromagnetic transition with moments ordered along the easy $c$-axis.

At higher temperatures, the data above 100 K can be analyzed using the Curie-Weiss law: $\chi$=$\chi_0$+$C$/$(T-\theta_{\rm CW})$, where $\chi_0$ is a temperature-independent term, $C$ is the Curie constant and $\theta_{\rm CW}$ is the Curie-Weiss temperature, yielding  $\theta^c_{\rm CW}=-11.7(3)$~K and an effective moment of  $\mu\rm_{eff}^c$ = 2.35$\mu_B$/Ce  for $H\parallel c$, as well as  $\theta^{ab}_{\rm CW}=-12.9(8)$~K and $\mu\rm_{eff}^{ab} = 2.49\mu_B$/Ce for $H \parallel ab$. The obtained values of $\mu\rm_{eff}$ for both directions are close to the full value of  $2.54~\mu_B$ for the $J=\frac{5}{2}$ ground state multiplet of Ce$^{3+}$. At lower temperatures, there is a deviation of  $\chi(T)$ from Curie-Weiss behavior, due to the splitting of the ground state multiplet by crystalline-electric fields. To analyze the CEF level scheme, we considered the following Hamiltonian for a  Ce$^{3+}$ ion in a tetragonal CEF \cite{hutchings1964point}

\begin{equation}
\begin{aligned}
\mathcal{H}_{\rm CF} =B_{2}^{0} O_{2}^{0}+B_{4}^{0} O_{4}^{0}+B_{4}^{4} O_{4}^{4}
\end{aligned}
\end{equation}
\noindent where $O_{l}^{m}$ and $B_{l}^{m}$ are Stevens operator equivalents and  parameters, respectively. The $B_2^0$ parameter can be estimated from the high temperature susceptibility using \cite{JensenBook}

 \begin{equation}
\begin{aligned}
B_{2}^{0}=\frac{10 k_{B}\left(\theta^{ab}_{\rm CW}-\theta^{c}_{\rm CW}\right)}{3(2 J-1)(2 J+3)},
\end{aligned}
\label{B20eq}
\end{equation}

\noindent where $J=\frac{5}{2}$ for the ground state multiplet of Ce$^{3+}$, yielding $B^0_2$ = -0.01077 meV. $\chi(T)$ along both directions was analyzed taking into account the contribution from the CEF $\chi^{i}_{\rm CEF}$, as well as molecular field parameters $\lambda^{i}$ using

 \begin{equation}
\begin{aligned}
\chi^i=\chi^{i}_{0}+\frac{\chi^{i}_{\mathrm{CEF}}}{1-\lambda^{i} \chi^{i}_{\mathrm{CEF}}},
\end{aligned}
\end{equation}

\noindent where the superscript $i$ denotes the $c$-axis or $ab$-plane. With $B^0_2$ fixed from Eq.~\ref{B20eq}, values of  $B^0_4$ = -0.0746 meV and $|B^4_4|$ = 0.496 meV were obtained, together with molecular field parameters of $\lambda^c$ = -3.55~mol/emu and   $\lambda^{ab}$ = 8.15~mol/emu, $\chi^{c}_{0}=2.2\times10^{-4}$emu/mol and $\chi^{ab}_{0}=-2.3\times10^{-3}$emu/mol, and the fitted results are shown in Fig.~\ref{fig4}(b). These parameters yield a CEF scheme with a $\Gamma_7$ ground state Kramer’s doublet  $\left|\psi_{1}^{\pm}\right\rangle =0.883\left|\pm \frac{5}{2}\right\rangle-0.469\left|\mp \frac{3}{2}\right\rangle$ (for positive $B^4_4$), and excitations to $\Gamma_6$ and $\Gamma_7$ levels of $\Delta_1$ = 2.8 meV and  $\Delta_2$ = 32.1 meV, respectively. At high temperatures, the small negative $B^0_2$ leads to a nearly isotropic  $\chi(T)$, while at low temperatures, the negative $B^0_4$ leads to the observed Ising anisotropy with an easy $c$-axis. The predicted moment along the $c$-axis is given by $\left\langle\mu_{z}\right\rangle=\left\langle\psi_{1}^{\pm}\left|g_{J} J_{z}\right| \psi_{1}^{\pm}\right\rangle=1.4~\mu_B$/Ce, which is larger than the value obtained from the saturated magnetization. The positive value of $\lambda^{ab}$ is consistent with ferromagnetic coupling between spins within the basal plane, while the smaller negative $\lambda^c$ is consistent with weaker antiferromagnetic coupling between Ce layers.

\begin{figure}[t]
\includegraphics[width=8.6cm]{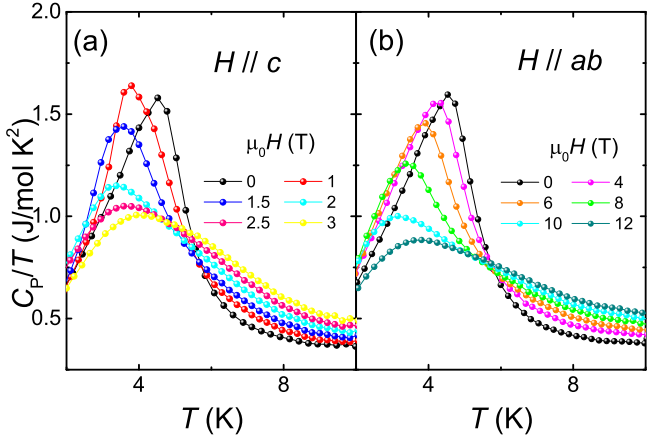}
\caption{(Color online) Temperature dependence of the specific heat of CePdGa$_{6}$ in various  applied magnetic fields (a) parallel to the $c$-axis, and (b) within the $ab$-plane.
 }
\label{fig5}
\end{figure}

\begin{figure}[t]
\includegraphics[width=8.6cm]{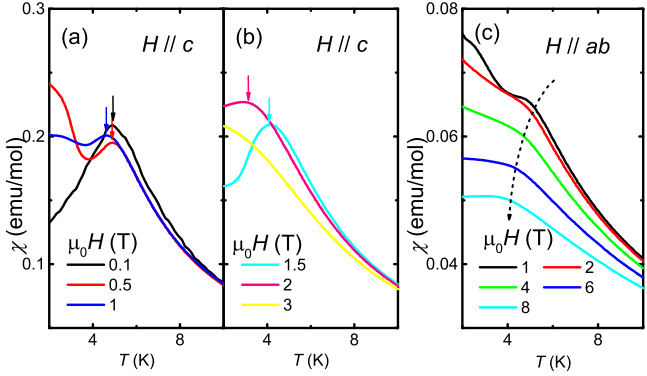}
\caption{(Color online) Temperature dependence of the magnetic susceptibility $\chi(T)$ of CePdGa$_{6}$ in different magnetic fields parallel to the $c$-axis for fields (a) below, and (b) above 1~T. The vertical arrows mark the position of the antiferromagnetic transition. Panel (c) shows $\chi(T)$ for various fields applied  within the $ab$-plane, where the dashed line shows the evolution of $T_{\rm N}$ with field.
}
\label{fig6}
\end{figure}

\begin{figure}[t]
\includegraphics[width=8.6cm]{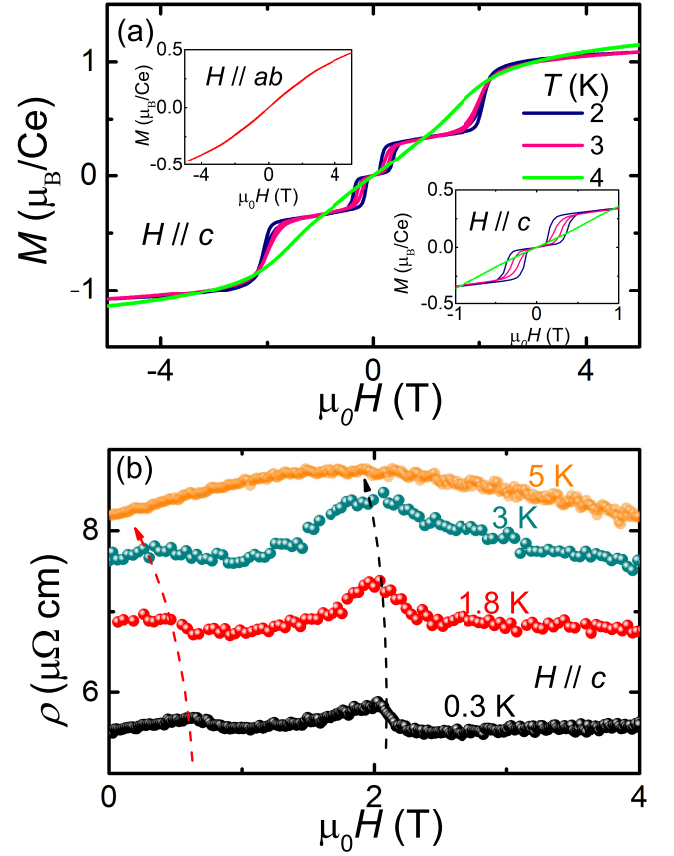}
\caption{(Color online) (a) Isothermal field dependence of the magnetization $M(H)$ of CePdGa$_{6}$ for fields along the $c$-axis, at three temperatures below $T_{\rm N}$. The lower inset displays the low field region of the data in the main panel, demonstrating hysteresis about the metamagnetic transition, while the upper inset shows $M(H)$ at 2~K for fields within the $ab$-plane. (b) Field dependence of the resistivity $\rho(H)$ of CePdGa$_6$ at several temperatures for fields along the $c$-axis. The dashed lines show the evolution of the two metamagnetic transitions.
}
\label{fig7}
\end{figure}

\begin{figure}[h]
\includegraphics[width=8.6cm]{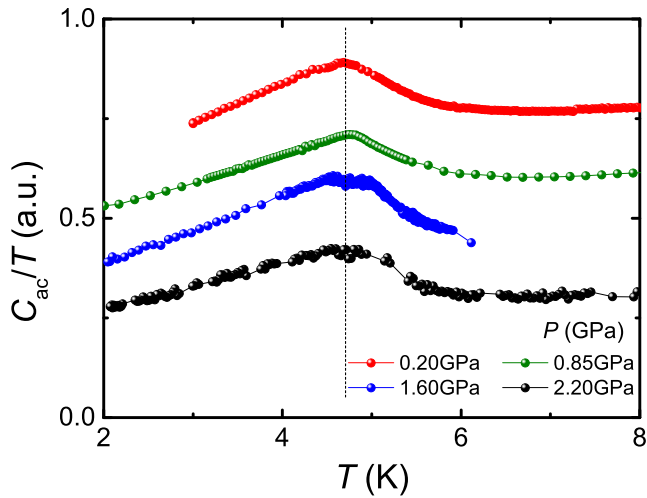}
\caption{(Color online) Temperature dependence of the ac heat capacity of CePdGa$_6$ at various hydrostatic pressures up to 2.2~GPa. The vertical dashed line shows the position of the ambient pressure $T_{\rm N}$, which remains nearly unchanged with pressure.
 }
\label{fig8}
\end{figure}

\begin{figure}[h]
\includegraphics[width=8.6cm]{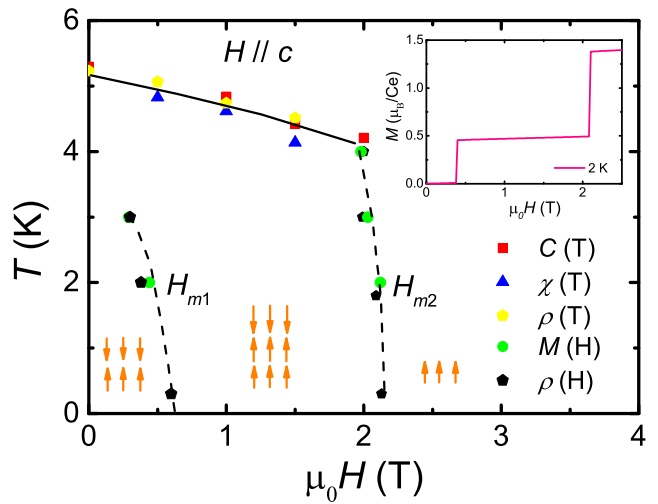}
\caption{(Color online) Temperature-field phase diagram of CePdGa$_6$ at ambient pressure for fields along the easy $c$-axis, from measurements of the resistivity, magnetization, and specific heat. The solid line shows the evolution of $T_{\rm N}$, while the dashed lines show the positions of the low temperature metamagnetic transitions. The magnetic structures at low temperature are also illustrated by the orange arrows, where in zero-field there is an antiferromagnetic ground state, while upon applying a field the system passes through an intermediate $\uparrow\uparrow\downarrow$ phase, before entering the spin polarized state. The inset shows the field dependence of the magnetization based on mean-field calculations of the magnetic ground state calculated using the \textsc{McPhase} software package \cite{Rotter2004}, with the parameters described in the text.
 }
\label{fig9}
\end{figure}

\subsection{Field dependence of the magnetic properties}

In order to determine the behavior of the magnetic ground state in magnetic fields, and to map the field-temperature phase diagrams, measurements of the specific heat and magnetization were performed in different applied fields. Figure~\ref{fig5}(a) displays the low temperature specific heat of CePdGa$_{6}$ with different fields applied along the $c$-axis. It can be seen that $T\rm_{N}$ is gradually suppressed with increasing field, and at fields greater than 2~T, no magnetic transition is observed. Instead, there is a broad hump in $C/T$, which shifts to higher temperature with increasing field, corresponding to the Schottky anomaly from the splitting of the ground state doublet in the applied field. In Fig.~\ref{fig5}(b), $C/T$ is displayed for fields within the $ab$-plane, where the antiferromagnetic transition is more robust than for fields along the $c$-axis, and the broad Schottky anomaly is only clearly resolved in a field of 12~T. The differences in the field dependence for the two different field directions is consistent with the low temperature Ising anisotropy in CePdGa$_{6}$, where a smaller field along the easy $c$-axis can bring the system to the spin-polarized state.

The low temperature $\chi(T)$ in different applied fields are displayed in Fig.~\ref{fig6}. For fields along the $c$-axis distinctly different behaviors are observed for different field ranges. In a field of 0.1~T, there is a sharp peak at $T\rm_{N}$, corresponding to entering the antiferromagnetic ground state. At a larger field of 0.5~T, only a small hump is observed at $T\rm_{N}$, while at low temperatures there is an increase in $\chi(T)$, and at higher fields there is broad peak which is gradually suppressed with field. Meanwhile for fields within the $ab$-plane up to at least 8~T, there is a gradual suppression of $T\rm_N$, in line with the specific heat results.

The isothermal magnetization as a function of field along the $c$-axis  at three temperatures below $T\rm_{N}$ is displayed in Fig.~\ref{fig7}(a), measured upon both sweeping the field up and down. In zero-field there is no remanent magnetization, consistent with a purely antiferromagnetic ground state. At 2~K, there are two metamagnetic transitions  at $H_{m1}=0.4$~T and $H_{m2}=2.1$~T, where hysteresis is also observed indicating a first-order nature, whereas otherwise the magnetization plateaus, with only a weak change of the magnetization with field. This is consistent with $H_{m1}$ and $H_{m2}$ corresponding to spin-flip transitions, with the spins remaining orientated along the $c$-axis. For fields above $H_{m2}$, no magnetic transition is observed in the specific heat, and therefore this likely corresponds to the system reaching the spin polarized state, with a saturation magnetization of $M_s=1.1~\mu_{\rm B}$/Ce. On the other hand, above $H_{m1}$ the magnetization reaches a value of $0.35~\mu_{\rm B}$/Ce, corresponding to $\approx M_s/3$, indicating a change of magnetic structure with a ferromagnetic component. While there is little change in the field-dependence of the magnetization at 3~K, the curves at 4~K are drastically different. Instead of there being abrupt step-like metamagnetic transitions, the magnetization smoothly increases with field, reaching a very similar saturation value. This suggests that at higher temperatures, the spins continuously rotate upon increasing the applied field, rather than undergoing abrupt spin flip transitions. The field dependent magnetization at 2~K for fields  in the $ab$-plane is also shown in the inset of Fig.~\ref{fig7}(a), which smoothly changes with field, with no sign of saturation up to at least 5~T, consistent with this being the hard direction of magnetization. The metamagnetic transitions are also revealed in the field dependence of the resistivity $\rho(H)$, as displayed in Fig.~\ref{fig7}(b) for fields along the $c$-axis. At 0.3~K, two abrupt anomalies are observed corresponding to $H_{m1}$ and $H_{m2}$, which are also detected at 1.8~K and 3~K. Above these transitions, there is a decrease of $\rho(H)$, consistent with the reduced spin-flip scattering arising from a larger ferromagnetic component to the magnetism. On the other hand, no metamagnetic transitions are detected at 5~K, where instead there is a broad peak in $\rho(H)$, again consistent with a more gradual reorientation of the spins with field at higher temperatures.

\subsection{Magnetism of CePdGa$_{6}$ under pressure }

To determine the evolution of the magnetic order under pressure, the temperature dependence of the ac specific heat of CePdGa$_{6}$ was measured at several different hydrostatic pressures up to 2.2~GPa, which are displayed in Fig.~\ref{fig8}. It can be seen from the dotted line that there is little change of $T\rm_N$ with pressure indicating the robustness of magnetic order. In the case of the layered  Ce$_2M$Ga$_{12}$ compounds, the $T\rm_N$ of Ce$_2$NiGa$_{12}$ and Ce$_2$PdGa$_{12}$ decrease with pressure, and antiferromagnetism is suppressed entirely above 5.5 and 7~ GPa, respectively \cite{Kawamura_2014,Ohara_2012}. On the other hand the $T\rm_N$ of Ce$_2$IrGa$_{12}$ undergoes a moderate enhancement from  3.1 to 3.7~K for pressures up to 2.3~GPa, indicating that this compound is located on the left side of the Doniach phase diagram \cite{PhysRevB.101.024421}. In the case of CePdGa$_6$, the robustness of $T\rm_N$ suggests that measurements to higher pressures are required to situate this compound within the framework of the Doniach phase diagram and to examine whether there is pressure-induced quantum criticality in CePdGa$_{6}$.

\section{DISCUSSION}

Our measurements of the resistivity, magnetic susceptibility and specific heat  show that CePdGa$_{6}$ orders antiferromagnetically below $T\rm_N$ = 5.2~K, with the moments orientated along the $c$-axis. Figure~\ref{fig9} displays the temperature-field phase diagram for magnetic fields applied along the $c$-axis. The phase boundaries obtained from different measurements are highly consistent, showing that $T\rm_N$ shifts to lower temperatures with field, before abruptly disappearing in a field of 2~T. At  low temperatures, there are two step-like metamagnetic transitions shown by the dashed lines, where the second transition is to the spin polarized state, while the lower transition corresponds to a change of magnetic state to a phase with a magnetization of 0.35 $\mu\rm_{B}$/Ce, about one-third of the saturated value. Such step-like changes in the magnetization suggest that the spins are strongly constrained along the $c$-axis, and therefore there are abrupt spin-flip transitions for fields applied along the ordering direction. On the other hand, at 4~K the magnetization changes smoothly with field, reaching the same saturated magnetization, indicating that at this temperature the spins continuously rotate in the applied field. Such a change with temperature may be a consequence of only a moderate magnetocrystalline anisotropy, as also evidenced by the relatively small value of the spin-wave gap $\Delta_{\mathrm SW}/T_{\rm N}\approx0.4$, as compared to the other heavy fermion gallides Ce$_2$IrGa$_{12}$ and Ce$_2$PdGa$_{12}$ which have $\Delta_{\mathrm SW}/T_{\rm N}$ of 1.5 and 2.8, respectively \cite{PhysRevB.101.024421,Gnida_2013}.

From the analysis of the magnetic susceptibility including the CEF contribution, the molecular field parameter is positive in the $ab$-plane ($\lambda^{ab}$), while a smaller negative value is obtained along the $c$-axis ($\lambda^{c}$). Together with the fact that only a relatively small field along the $c$-axis is required to reach the spin polarized state, this suggests that the antiferromagnetic ground state consists of ferromagnetically ordered Ce-layers coupled antiferromagnetically along the $c$-axis. The simplest model for such a system would consist of ferromagnetic Heisenberg exchange interactions between nearest neighbor Ce atoms within the $ab$-plane $\mathcal{J}_0>0$, and antiferromagnetic exchange interactions $\mathcal{J}_1<0$ between nearest neighboring layers, as well as a sufficiently strong  Ising anisotropy. This yields an $A$-type antiferromagnetic  ground state consisting of ferromagnetic layers with moments orientated along the $c$-axis, where the moment direction alternates between adjacent layers, ``$\uparrow\downarrow\uparrow\downarrow$''. This model however cannot account for the field induced phase with one-third magnetization, since for fields along the $c$-axis, only a metamagnetic transition directly from the $\uparrow\downarrow\uparrow\downarrow$ phase to the spin polarized state is anticipated.

In order to realize the intermediate field-induced phase, it is necessary to consider an antiferromagnetic exchange $\mathcal{J}_2$ between next nearest neighboring layers. In this case, from considering the classical ground state energies with sufficiently strong Ising anisotropy, the same $\uparrow\downarrow\uparrow\downarrow$ ground state is realized for $\mathcal{J}_1/\mathcal{J}_2>2$, while a $\uparrow\uparrow\downarrow\downarrow$ state occurs for $\mathcal{J}_1/\mathcal{J}_2<2$ \cite{Li2019}. Upon applying a magnetic field along the $c$-axis, there is a metamagnetic transition at a field $H_{m1}$ to an $\uparrow\uparrow\downarrow$ state with a net magnetization one-third of the saturated value, and another at $H_{m2}$ to the spin polarized state, where  $H_{m2}/H_{m1}$ is determined by $\mathcal{J}_1/\mathcal{J}_2$. We performed mean-field calculations of the magnetic ground state and magnetization using the \textsc{McPhase} software package \cite{Rotter2004}, which determines the most stable magnetic structure at a given temperature and magnetic field from considering  multiple random starting moment configurations. These took into account the Heisenberg exchange interactions described above, as well as the CEF Hamiltonian $\mathcal{H}_{\rm CF}$ with our deduced values of the Stevens parameters. As shown in the inset of Fig.~\ref{fig9},  the observed values of $H_{m1}=0.4$~T and $H_{m2}=2.1$~T, from the midpoints of the metamagnetic transitions at 2~K, are well reproduced from the mean-field calculations at 2~K with $\mathcal{J}_1=-0.023$~meV and $\mathcal{J}_2=-0.0085$~meV, where for $H_{m1} < H < H_{m2}$ the $\uparrow\uparrow\downarrow$ ground state has the lowest energy. Keeping these values fixed, we find that a nearest neighbor in-plane ferromagnetic interaction $\mathcal{J}_0=0.034$~meV can yield the observed value of $T_{\rm N}=5.2$~K. Therefore our analysis suggests stronger in-plane ferromagnetic interactions, where the value of  $4\mathcal{J}_0/(2\mathcal{J}_1+2\mathcal{J}_2)=2.16$ is close to our fitted value of $\lambda^{ab}/\lambda^c=2.3$. Note that here we have assumed a $\uparrow\downarrow\uparrow\downarrow$ ground state with $\mathcal{J}_1/\mathcal{J}_2>2$. Although a $\uparrow\uparrow\downarrow\downarrow$ phase has been reported in  CeCoGe$_3$ \cite{PhysRevB.88.134416}, such a scenario is less likely in CePdGa$_6$ due to the larger interlayer distances.

Compared to the layered heavy fermion antiferromagnet CeRhIn$_5$, the magnetism in CePdGa$_6$ appears to have a much more three dimensional character, whereas it is rather two-dimensional in the former, with $\mathcal{J}_1/\mathcal{J}_0=0.13$ deduced from inelastic neutron scattering \cite{Das2014}. In addition, in CeRhIn$_5$ the easy plane anisotropy and presence of in-plane antiferromagnetic interactions give rise to spiral magnetic order which is incommensurate along the $c$-axis \cite{Curro2000,Bao2000}, and these features may be important factors for realizing the unconventional quantum criticality and superconductivity. On the other hand, the $T_{\rm N}$ of CePdGa$_6$ is much more robust with pressure, remaining almost unchanged at pressures  up to 2.2~GPa. Therefore an understanding of the relationship between the magnetism and any quantum critical behaviors will require measurements at considerably higher pressures.

In addition, despite the layered arrangement of Ce atoms, the \textit{local} environment of the Ce atoms is relatively three dimensional, as evidenced by the derived CEF parameters being close to that for a cubic system (where $B_2^0=0$ and $|B_4^4|=5|B_4^0|$). This CEF scheme can correctly predict the low-temperature Ising anisotropy, but the predicted moment along the $c$-axis is larger than that observed. While such a reduced moment compared to that predicted from the  CEF level-scheme is often observed in heavy fermion  antiferromagnets due to screening of the moments by the Kondo effect \cite{PhysRevB.88.134416,Bao2000,Christianson2002,Stockert2004,Goremychkin1993}, confirming whether such a scenario is applicable to CePdGa$_6$ requires a more precise determination of the CEF parameters, by measurements such as inelastic neutron scattering.

\section{Conclusion}

In summary, we have characterized the magnetic properties of the heavy fermion antiferromagnet CePdGa$_6$, and their evolution upon the application of magnetic fields and pressure. We have constructed the temperature-field phase diagram for fields along the $c$-axis, where at low temperatures there are two abrupt metamagnetic transitions corresponding to spin-flip transitions. From the analysis of the magnetic susceptibility, we propose a CEF level scheme for the splitting of the ground state $J=5/2$ multiplet, indicating that the Ising anisotropy at low temperatures is driven by the sizeable $B_4^0$ parameter. Moreover, our results are consistent with an antiferromagnetic ground state consisting of ferromagnetically coupled Ce-layers, with antiferromagnetic coupling between layers. We have proposed a model for the exchange interactions which can explain the evolution of the magnetic ordering with applied magnetic field, which has sizeable nearest neighbor and next-nearest neighbor layer interactions, indicating the presence of significant long-range magnetic interactions. Despite evidence for heavy fermion behavior, there is negligible change of $T_{\rm N}$ upon applying pressures up 2.2~GPa, and hence measurements at much higher pressures are necessary to look for evidence of quantum criticality.

\section{acknowledgments}

We are grateful to Martin Rotter for advice with the \textsc{McPhase} software. This work was supported by the National Key R$\&$D Program of China (2017YFA0303100), the Key R$\&$D Program of Zhejiang Province, China (2021C01002), and the National Natural Science Foundation of China (12174332, 12034017 and 11974306).

%

\end{document}